\documentclass{aa}

\usepackage{amsmath}
\usepackage{graphicx}
\usepackage{natbib}
\usepackage{txfonts}
\usepackage{ulem}
      \def\new#1 {{\bf #1 }}
      \def\cut#1 {\sout{#1} }

\def\thc{$^{13}$C}

\def\dcop{DCO$^+$}

\def\ammo{NH$_3$}

\def\hhdp{H$_2$D$^+$}

\def\nnhp{N$_2$H$^+$}
\def\nndp{N$_2$D$^+$}

\def\hcccn{HC$_3$N}

\def\vlsr{$V_{\rm LSR}$}

\def\kms{km~s$^{-1}$}

\def\tmb{$T_{\rm MB}$}

\def\aap{{A\&A}}

\def\apjl{{ApJ}}
\def\apjs{{ApJS}}

\begin{document}

\title{Hyperfine structure in the $J = 1 - 0$ transitions of DCO$^+$, DNC, and 
       HN$^{13}$C: astronomical observations and quantum-chemical calculations}

\author{Floris F.~S.\ van der Tak\inst{1,3} 
        \and Holger S.~P.\ M\"uller\inst{2,3}
        \and Michael E. Harding\inst{4,5}
        \and J\"urgen Gauss\inst{4}}

      \institute{ SRON Netherlands Institute for Space Research, Landleven 12,
        9747 AD Groningen, The Netherlands \\
        \email{vdtak@sron.nl} 
        \and 
        I.\ Physikalisches Institut der Universit\"at, Z\"ulpicher Stra{\ss}e 77, 
        50937 K\"oln, Germany 
        \and
        Max-Planck-Institut f\"ur Radioastronomie, Auf dem H\"ugel 69, 
        53121 Bonn, Germany 
        \and 
        Institut f\"ur Physikalische Chemie, Universit\"at Mainz, 55099 Mainz, Germany
        \and 
        Department of Chemistry and Biochemistry,
        University of Texas, Austin, TX 78712, U.S.A.}

\titlerunning{Hyperfine structure of DCO$^+$, DNC and HN\thc }
\authorrunning{Van der Tak et al.}

\date{Received 17 July 2009 / Accepted 27 August 2009}

\abstract
% context heading (optional)
{Knowledge of the hyperfine structure of molecular lines is useful for
  estimating reliable column densities from observed emission, and essential for the
  derivation of kinematic information from line profiles.}
% aims heading (mandatory)
{Deuterium bearing molecules are especially useful in this regard, because they
  are good probes of the physical and chemical structure of molecular cloud
  cores on the verge of star formation. However, the necessary spectroscopic
  data are often missing, especially for molecules which are too unstable for
  laboratory study.}
% methods heading (mandatory)
{ We have observed the ground-state ($J = 1 - 0$) rotational transitions of DCO$^+$, 
  HN\thc\ and DNC with the IRAM 30m telescope toward the dark cloud LDN 1512 
  which has exceptionally narrow lines permitting hyperfine splitting to be resolved
  in part.  The measured splittings of 50--300\,kHz are used to derive nuclear 
  quadrupole and spin-rotation parameters for these species. The measurements 
  are supplemented by high-level quantum-chemical calculations using 
  coupled-cluster techniques and large atomic-orbital basis sets. }
% results heading (mandatory)
{ We find $eQq = + 151.12~(400)$~kHz and $C_I = -1.12~(43)$~kHz for \dcop,
  $eQq = 272.5~(51)$~kHz for HN\thc, and $eQq({\rm D})=265.9(83)$\,kHz and
  $eQq({\rm N}) = 288.2~(71)$\,kHz for DNC. The numbers for DNC are consistent
  with previous laboratory data, while our constants for \dcop\ are somewhat
  smaller than previous results based on astronomical data. For both \dcop\ and
  DNC, our results are more accurate than previous determinations. Our results
  are in good agreement with the corresponding best theoretical estimates, which 
  amount to $eQq = 156.0$~kHz and $C_I = -0.69$~kHz for \dcop, $eQq = 279.5$~kHz 
  for HN\thc, and $eQq({\rm D}) = 257.6$~kHz and $eQq({\rm N}) = 309.6$~kHz for
  DNC. We also derive updated rotational constants for HN\thc:
  $B$=43545.6000(47)\,MHz and $D$=93.7(20)\,kHz.}
% conclusions heading (optional)
{ The hyperfine splittings of the \dcop, DNC and HN\thc\ $J = 1 - 0$ lines range
  over 0.47--1.28\,\kms, which is comparable to typical line widths in
  pre-stellar cores and to systematic gas motions on $\sim$1000\,AU scales in
  protostellar cores. We present tabular information to allow inclusion of the
  hyperfine splitting in astronomical data interpretation.  The large
  differences in the $^{14}$N quadrupole parameters of DNC and HN$^{13}$C have
  been traced to differences in the vibrational corrections caused by
  significant non-rigidity of these molecules, particularly along the bending
  coordinate.}


\maketitle

%%%%%%%%%%%%%%%%%%%%%%%%%%%%%%%%%%%%%%%%%%%%%%%%%%%%%%%%%%%%%%%%%%%%%%%%%%%%%%
\section{Introduction}
\label{s:intro}
%%%%%%%%%%%%%%%%%%%%%%%%%%%%%%%%%%%%%%%%%%%%%%%%%%%%%%%%%%%%%%%%%%%%%%%%%%%%%%

Cold interstellar clouds have long been recognized as excellent laboratories for
determining basic physical quantities of molecular structure (for a review see
\citealt{mol-spa-lab}).  In particular, these clouds provide access to molecular
species that are too unstable to permit sufficient terrestrial production for
in-depth investigation. Some molecules were even detected in interstellar space
before they were found on Earth. A classic example is the X-ogen of the early
1970's, which has gained prime importance for studying the interactions between
interstellar gas and magnetic fields since its identification as HCO$^+$
\citep{buhl:x-ogen,kraemer:hco+}.

Astronomical observations of molecules are not only useful to provide accurate
rest frequencies of spectral lines \citep{pagani:c34s,pagani:freqs}, but also to
determine hyperfine parameters.  The best-known case of hyperfine splitting in
molecules with no unpaired electrons is the electric quadrupole splitting which
occurs for nuclei with spin $I \ge 1$ such as, for example, D and $^{14}$N. A
second type of splitting occurs if the molecule contains nuclei with $I > 0$
such as H, $^{13}$C, and $^{15}$N due to magnetic spin-rotation coupling and/or
nuclear spin-nuclear spin coupling.
Both effects are usually significantly smaller than the splitting due to
electron spin-nuclear spin coupling in molecules with an unpaired electron and
an $I > 0$ nucleus.

A recent example of astronomically determined hyperfine parameters is the
determination of spectroscopic parameters of H$^{13}$CO$^+$ based on mm-wave
observations of the dark cloud LDN 1512 \citep{jsb04}. The exceptionally narrow
lines in this cloud ($\Delta V = 0.16$~\kms) allow a frequency resolution and
accuracy that usually cannot be attained in the laboratory for short-lived
molecules. This category includes molecular ions, but also radicals and reactive
species such as DNC, which has a lifetime of $< 1$~s in the
laboratory. Schmid-Burgk et al. showed that even the unresolved hyperfine
splitting of $^{13}$CO plays a role in the case of very narrow lines.
They used their spectra of LDN~1512 not only to determine the spin-rotation
constant $C_I$ of the $^{13}$C nucleus in H$^{13}$CO$^+$, but also 
to show that the magnitude of the splitting and the intensity ratio of the 
two resolved features depends on a much smaller effect, namely $C_I$ of 
the H nucleus and the nuclear spin-nuclear spin coupling between 
these two nuclei \citep{jsb04}.

In astrophysics, the prime use of hyperfine splitting is the possibility of
measuring the optical depths of molecular lines.  This fundamental quantity
allows estimation of molecular column densities without assumptions about the
beam filling factor. The classic example is the splitting of the \ammo\
inversion lines \citep{ho:ammo} which are used widely to measure the kinetic
temperatures of dense interstellar clouds \citep{walmsley:nh3}.
In addition, knowledge of hyperfine structure is essential for deriving the
kinematic structure of clouds from observations of molecular line profiles. In
particular, the central regions of pre-stellar cores are currently of great
interest, as the places where the transition from spherical infall to disk-like
rotation occurs \citep{bergin:araa}.  However, in these objects, many 'standard'
kinematic probes are unavailable due to freeze-out onto dust grains
\citep{bergin}. Deuterium bearing molecules are abundant even under these
conditions, but their lines always exhibit hyperfine splitting due to the
nonzero spin ($I = 1$) of the D nucleus. For example, \citet{vdtak:l1544} used
observations of \hhdp\ to study the kinematics at the center of the pre-stellar
core LDN 1544, where most other molecules are frozen onto dust grains.

In the laboratory, the Doppler-limited line widths in the 3\,mm region are of
the order of 200\,kHz, making it possible to resolve moderate to large hyperfine
splitting.  However, very small quadrupole splitting, such as the $^{14}$N
splitting in HNC or essentially all of the deuterium quadrupole splitting cannot
be resolved in this frequency region.  Sub-Doppler resolution techniques, such
as Lamb-dip spectroscopy or molecular beam millimeter-wave Fourier transform
spectroscopy, are sometimes available, but the former is usually not feasible for 
short-lived species, and the latter type of measurements is available only in 
very few laboratories.

While the relatively large nitrogen quadrupole splitting of DCN has been
resolved in the laboratory \citep{Lamb-dip_DCN} and in space \citep{turner}, the
hyperfine structure of the astrophysically important \dcop, HN\thc\ and
DNC species had not yet been resolved at the time of our observations
  (2004-2005).  High-resolution
spectroscopy of these species is not only astronomically useful to determine the
optical depths of the lines, but also to verify the results of quantum-chemical
calculations of spectroscopic constants.  For example, \citet{frerking} measured
a nitrogen quadrupole moment of 0.28~(3)~MHz for HNC in the dark cloud
LDN~134N. Even though this value is not particularly accurate, it is in good
agreement with the very recent, more accurate laboratory value of
0.2645~(46)~MHz \citep{bechtel:hnc-dnc}.  More recently, \citet{caselli:n2h+}
determined hyperfine parameters for the \nnhp\ $J = 1 - 0$ line from
astronomical data, which \citet{gerin:n2h+} extended to higher-$J$ transitions,
and \citet{dore04} to the \nndp\ isotopologue.  This paper describes a
new determination of the hyperfine structure of the $J = 1 - 0$ lines of \dcop,
DNC and HN\thc, based on observations of LDN 1512.  The measurements are
supplemented by high-level quantum-chemical calculations of the corresponding
hyperfine parameters using state-of-the-art coupled-cluster techniques together
with large atomic-orbital basis sets.

%%%%%%%%%%%%%%%%%%%%%%%%%%%%%%%%%%%%%%%%%%%%%%%%%%%%%%%%%%%%%%%%%%%%%%%%%%%%%%
\section{Observations}
\label{s:obs}
%%%%%%%%%%%%%%%%%%%%%%%%%%%%%%%%%%%%%%%%%%%%%%%%%%%%%%%%%%%%%%%%%%%%%%%%%%%%%%

Observations of the $J = 1 - 0$ lines of \dcop, DNC and HN\thc\ near 72039,
76306 and 87091 MHz were performed on 2004 August 5 -- 6, and 2005 March 12 and
July 29.  The 30-m telescope of the Institut de Radio Astronomie Millim\'etrique
(IRAM)\footnote{IRAM is an international institute for research in millimeter
  astronomy, co-funded by the Centre National de la Recherche Scientifique
  (France), the Max Planck Gesellschaft (Germany) and the Instituto Geografico
  Nacional (Spain).} was used, with the facility receivers A100 and B100 as
front end.  The tuning range of these receivers had just been extended from
80\,GHz down to 70~GHz.  The beam size is 33$''$ in this frequency range, and
the main beam efficiency is $\approx$80\%.  We used the Versatile Spectral
Assembly (VESPA) correlator as backend to achieve a spectral resolution of
3.3~kHz, or 0.013~\kms.  Sideband gain ratios, measured with the Martin-Puplett
interferometer, were in the range 0.91--1.13, depending on backend module.
System temperatures were $\approx$300~K at 72~GHz, $\approx$200~K at 76~GHz and
$\approx$150~K at 87~GHz.  Integration times were 78 minutes (on+off) at 72~GHz,
215 minutes at 76~GHz and 197 minutes at 87~GHz, giving rms noise levels of
\tmb=158~mK at 72~GHz, 47~mK at 76~GHz and 41~mK at 87~GHz. Pointing was checked
every hour on the nearby planet Venus.

The position observed is $\alpha$ = 05:00:54.40, $\delta$ = $+$32:39:37.0
(B1950). The cloud velocity at this position, as determined from observations of
the HC$_3$N 3--2 line near 27294 MHz with the Effelsberg 100-m
telescope\footnote{The Effelsberg telescope is operated by the
  Max-Planck-Institut f\"ur Radioastronomie}, is \vlsr = 7.069$\pm$\,0.001~\kms.
The beam size of the 100-m telescope at the HC$_3$N frequency is almost equal to
that of the 30-m telescope at the \dcop, DCN and DNC frequencies, so the effect
of the known velocity gradients within the LDN 1512 cloud should be very small.
An upper limit is obtained by convolving the \hcccn\ data to a 50$''$ beam,
which gives a velocity of \vlsr\ = 7.063\,$\pm$\,0.001~\kms. Therefore the
effect of the velocity gradient in LDN 1512 is less than 0.006~\kms.  For
further discussion of these velocity gradients and for accurate \hcccn\
frequencies, see \citet{jsb04}.

The data reduction was performed using the Continuum and Line Analysis
Single-dish Software (CLASS) package\footnote{http://www.iram.fr/IRAMFR/GILDAS}
and followed standard procedures, except that the combination of data taken on
different epochs posed unusual challenges.  The telescope system corrects the
frequency scale of the spectra for heliocentric motion of the Earth, but the
correction is only exact for the center of the receiver bandpass. In our case,
the HN\thc\ line was observed at an offset of 160\,MHz to allow simultaneous
observation of other lines. The telescope system adds this offset as an absolute
value, but the orbital motion of the Earth actually dilates or shrinks the
frequency scale by a factor of $\sim$10$^{-4}$, the ratio of the telluric
orbital velocity and the velocity of light. The HN\thc\ spectra were taken 121
days apart, and the frequency shift is 16\,kHz $\times$ cos\,$\delta$, where
$\delta$ is the ecliptic latitude of the source (10$^\circ$). Indeed, the two
HN\thc\ spectra agree very well after a shift of 21\,kHz, and similarly, the DNC
spectra are consistent with the expected shift of 4.4\,kHz for an epoch
separation of 340 days. To obtain the correct absolute frequency scale, the
spectra of all epochs were shifted to a hypothetical observing date of June 8,
which is the day in the year that the Sun and LDN 1512 have the same hour angle
and transit at the same time.  We assume the frequency accuracy to be
3~kHz. This is about twice the LSR velocity gradient of the source and leaves
some room for possible errors in the correction of the heliocentric motion.

%%%%%%%%%%%%%%%%%%%%%%%%%%%%%%%%%%%%%%%%%%%%%%%%%%%%%%%%%%%%%%%%%%%%%%%%%%%%%%
\section{Quantum-chemical calculations}
\label{s:qcc}
%%%%%%%%%%%%%%%%%%%%%%%%%%%%%%%%%%%%%%%%%%%%%%%%%%%%%%%%%%%%%%%%%%%%%%%%%%%%%%

High-level quantum-chemical calculations using coupled-cluster (CC) techniques
\citep{cizek2,cizek1,review1,review2,review3} have been performed for the
hyperfine parameters of \dcop, HN\thc, and DNC. Calculations have been carried
out for the most part at the CC singles and doubles (CCSD) level \citep{CCSD}
augmented by a perturbative treatment of triple excitations 
(CCSD(T); \citealt{ccsdt}) which has proven in many cases to provide a reliable 
account of electron-correlation effects on energies and properties. 
The required one-particle basis sets for these calculations have been taken 
from Dunning's hierarchy of correlation-consistent basis sets with cc-pVXZ 
denoting the standard valence sets \citep{dunn1}, cc-pCVXZ those with additional 
core-polarizing functions \citep{dunn2}, and aug-cc-p(C)VXZ those 
with additional diffuse functions \citep{dunn3}. 
Here, X represents the cardinal number of the basis sets and values 
of 3 (= T), 4 (= Q), and 5 have been chosen in the present work.

The theoretical determination of quadrupole coupling constants is based on the
evaluation of the electric-field gradients (efgs) at the corresponding nuclei.
The nuclear quadrupole moments required to convert the efgs to the quadrupole
couplings are taken from the literature \citep{pekka}, and the following values
have been here adopted: 2.860~(15)~mbarn for D and 20.44~(3)~mbarn for $^{14}$N.
The reported calculations also take into account zero-point vibrational effects
on the quadrupole couplings. Those are treated in a perturbative manner as
described in \citet{auer} and necessitate the evaluation of quadratic and cubic
force fields. The latter are obtained using analytic second-derivative
techniques \citep{secder} as described in \citet{irpc}.  \citet{DC3N_ai_exp}
provide recent examples of the very good agreement between experiment and
calculations for DC$_3$N and HC$_3$N.

The nuclear spin-rotation tensor is a second-order response property and can be
computed via the associated second derivative of the energy with respect to
nuclear spin and the rotational angular momentum as perturbations. As described
in \citet{rlao}, we calculate the spin-rotation tensor using
perturbation-dependent basis functions (so-called rotational London atomic
orbitals) in order to ensure fast basis-set convergence. 
No vibrational averaging has been performed for the spin-rotation tensors, as the
corresponding theoretical expressions exhibit numerical problems in the case of 
polyatomic linear molecules.

To ensure convergence in the electron-correlation treatment, a few calculations have
been performed at levels beyond CCSD(T), i.e. at the CCSDT level \citep{ccsdft1,ccsdft2} 
with a full treatment of triple excitations and at the CCSDTQ level \citep{gcc1,gcc2} 
with an additional consideration of quadruple excitations.

All calculations have been carried out using the {\sc CFour} quantum-chemical
program package\footnote{Coupled Cluster techniques for Computational Chemistry,
  a quantum-chemical program package by J. F. Stanton, J.  Gauss, M. E. Harding,
  and P. G. Szalay with contributions from A. A.  Auer, R. J. Bartlett,
  U. Benedikt, D. E. Bernholdt, C. Berger, Y.J. Bomble, O.  Christiansen,
  M. Heckert, O. Heun, C. Huber, D. Jonsson, J.  Jus\'elius, K. Klein,
  W. J. Lauderdale, D. Matthews, T. Metzroth, D.  P. O'Neill, D. R. Price,
  E. Prochnow, K. Ruud, F. Schiffmann, S.  Stopkowicz, M. E. Varner,
  J. V\'azquez, J. D. Watts, F.  Wang and the integral packages MOLECULE
  (J. Alml\"of and P. R.  Taylor), PROPS (P. R. Taylor), ABACUS (T. Helgaker,
  H. J. Aa.  Jensen, P. J\o rgensen, and J. Olsen), and ECP routines by A. V.
  Mitin and C. van W\"ullen. For the current version, see
  \texttt{http://www.cfour.de}}.
Only the CCSDT and CCSDTQ calculations have been carried out with the {\sc MRCC}
package\footnote{MRCC, a string-based quantum chemical program suite written by
  M. K{\'a}llay.  See \texttt{http://www.mrcc.hu}} which has been interfaced to
{\sc CFour}.

Tables \ref{t:qcc-theor} and \ref{t:nsr-theor} summarize our computational
results for the hyperfine parameters of \dcop, HN\thc, and DCN and in 
particular document the convergence of the calculated values with respect
to basis set and electron-correlation treatment.

%%%%%%%%%%%%%%%%%%%%%%%%%%%%%%%%%%%%%%%%%%%%%%%%%%%%%%%%%%%%%%%%%%%%%%%%%%%%%%
\begin{table*}[p]
  \caption{Computed quadrupole-coupling constants $eQq$ (kHz) for \dcop,
HN\thc, and DNC.$^a$} 
\label{t:qcc-theor}

\begin{tabular}{lcccr@{}l}
\hline \hline
Computational level    & \multicolumn{1}{c}{\dcop} & \multicolumn{1}{c}{HN\thc} & \multicolumn{3}{c}{DNC} \\
\cline{4-6}
                       & D     & N     & D     & \multicolumn{2}{c}{N} \\
\hline
CCSD/cc-pCVDZ          &       &       & 308.6 & 1009&.5  \\
CCSD(T)/cc-pCVDZ       &       &       & 309.3 & 1021&.4 \\
CCSDT/cc-pCVDZ         &       &       & 309.3 & 1029&.1 \\ 
CCSDTQ/cc-pCVDZ        &       &       & 309.2 & 1033&.2 \\
CCSD(T)/cc-pCVTZ       &       &       & 290.8 &  431&.9 \\
CCSD(T)/cc-pCVQZ       & 159.2 & 341.6 & 279.8 &  341&.6 \\
CCSD(T)/cc-pCVQZ + vib & 153.6 & 294.8 & 259.1 &  325&.2 \\
CCSD(T)/cc-pCV5Z       & 162.2 & 325.5 & 278.3 &  325&.5 \\
CCSD(T)/cc-pCV5Z + vib & 156.0 & 279.5 & 257.6 &  309&.6 \\     
\hline
\end{tabular}

\medskip

$^a$ All calculations have been performed at geometries obtained at the
same computational level.

\end{table*}
%%%%%%%%%%%%%%%%%%%%%%%%%%%%%%%%%%%%%%%%%%%%%%%%%%%%%%%%%%%%%%%%%%%%%%%%%%%%%%

%%%%%%%%%%%%%%%%%%%%%%%%%%%%%%%%%%%%%%%%%%%%%%%%%%%%%%%%%%%%%%%%%%%%%%%%%%%%%%
\begin{table*}[p]
  \caption{Computed nuclear spin-rotation constants $C_I$ (kHz) for \dcop,
HN\thc, and DNC.$^a$} 
\label{t:nsr-theor}

\begin{tabular}{lclccclcc}
\hline \hline
Computational level & \multicolumn{1}{c}{\dcop} & & \multicolumn{3}{c}{HN\thc} & & \multicolumn{2}{c}{DNC} \\
\cline{4-6} \cline{8-9} 
                    & D        & & H        & N      & C     & & D        & N  \\
\hline
CCSD(T)/cc-pCVTZ    & $-$0.701 & & $-$4.551 & 22.954 & 6.194 & & $-$0.611 & 5.417 \\
CCSD(T)/cc-pCVQZ    & $-$0.695 & & $-$4.505 & 23.345 & 6.292 & & $-$0.605 & 5.502 \\
CCSD(T)/cc-pCV5Z    & $-$0.693 & & $-$4.486 & 23.471 & 6.330 & & $-$0.602 & 5.536 \\
CCSD(T)/aug-cc-pVTZ & $-$0.700 & & $-$4.575 & 22.878 & 6.200 & & $-$0.614 & 5.423 \\
CCSD(T)/aug-cc-pVQZ & $-$0.695 & & $-$4.510 & 23.318 & 6.295 & & $-$0.605 & 5.505 \\
CCSD(T)/aug-cc-pV5Z & $-$0.693 & & $-$4.486 & 23.456 & 6.331 & & $-$0.602 & 5.537 \\
\hline
\end{tabular}

\medskip

$^a$ The calculations have been performed at geometries obtained at the 
CCSD(T)/cc-pCVQZ level of theory 
(\dcop: $r({\rm CO})=1.10639$\,\AA, $r({\rm HC})=1.09236$\,\AA; HN\thc, 
DNC: $r({\rm NC})= 1.16927$\,\AA, $r({\rm HN})=0.99526$\,\AA).

\end{table*}
%%%%%%%%%%%%%%%%%%%%%%%%%%%%%%%%%%%%%%%%%%%%%%%%%%%%%%%%%%%%%%%%%%%%%%%%%%%%%%

%%%%%%%%%%%%%%%%%%%%%%%%%%%%%%%%%%%%%%%%%%%%%%%%%%%%%%%%%%%%%%%%%%%%%%%%%%%%%%
\section{Results}
\label{s:res}
%%%%%%%%%%%%%%%%%%%%%%%%%%%%%%%%%%%%%%%%%%%%%%%%%%%%%%%%%%%%%%%%%%%%%%%%%%%%%%

%%%%%%%%%%%%%%%%%%%%%%%%%%%%%%%%%%%%%%%%%%%%%%%%%%%%%%%%%%%%%%%%%%%%%%%%%%%%%%
\subsection{Hyperfine splitting of DCO$^+$}
\label{sec:dcop}
%%%%%%%%%%%%%%%%%%%%%%%%%%%%%%%%%%%%%%%%%%%%%%%%%%%%%%%%%%%%%%%%%%%%%%%%%%%%%%

%%%%%%%%%%%%%%%%%%%%%%%%%%%%%%%%%%%%%%%%%%%%%%%%%%%%%%%%%%%%%%%%%%%%%%%%%%%%%%
\begin{figure}[p]
\includegraphics[width=7cm,angle=0]{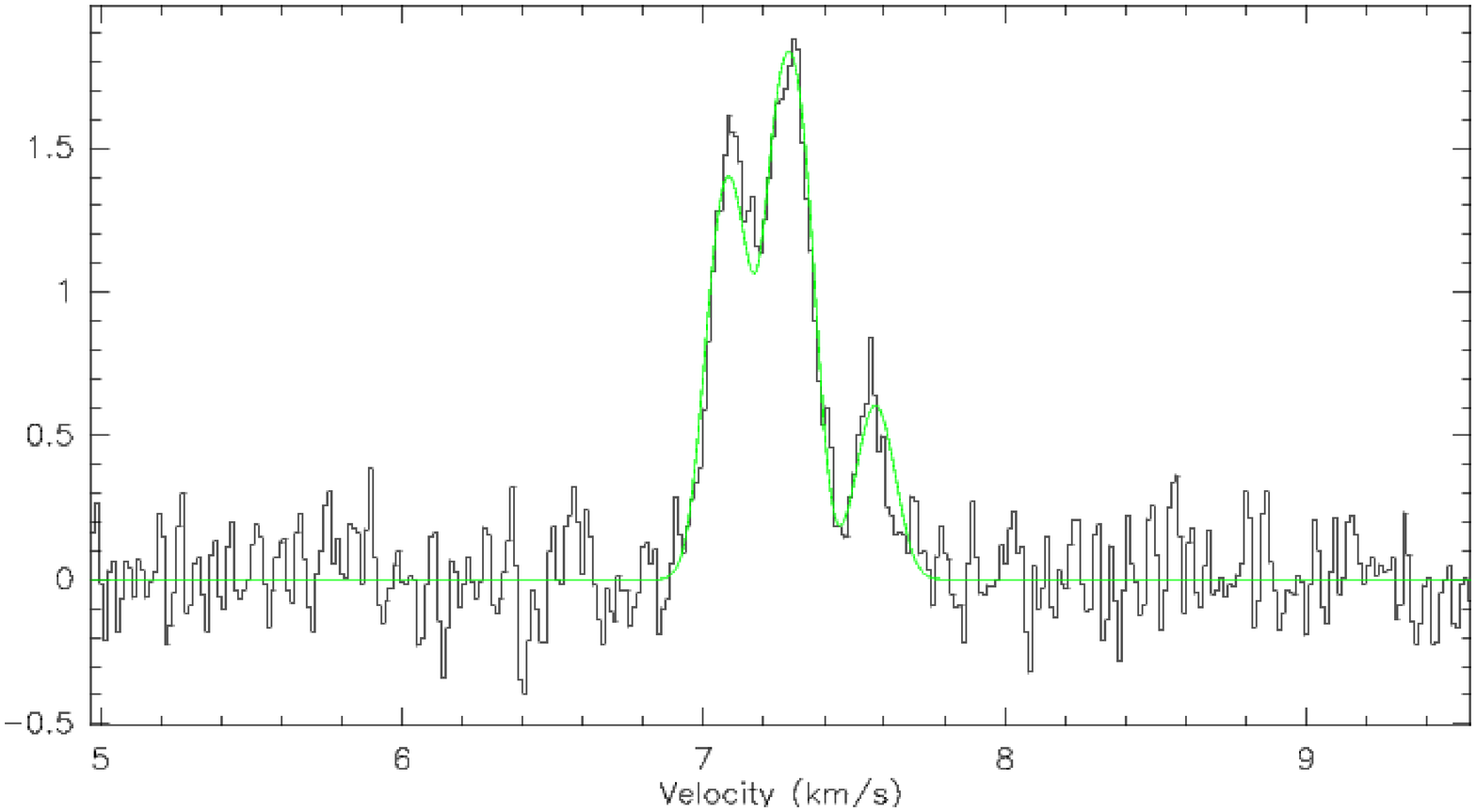}
\caption{Spectrum of the $J = 1 - 0$ transition of \dcop\ toward the dark cloud
  LDN 1512, obtained with the IRAM 30m telescope. Vertical scale is antenna
  temperature ($T_{\rm mb}$) in K. Overplotted is a model fit which assumes that
  the three hyperfine components have the same width and excitation
  temperature.}
\label{f:dcop}
\end{figure}
%%%%%%%%%%%%%%%%%%%%%%%%%%%%%%%%%%%%%%%%%%%%%%%%%%%%%%%%%%%%%%%%%%%%%%%%%%%%%%

Figure~\ref{f:dcop} shows the \dcop\ spectrum toward LDN 1512.  
The spin $I=1$ of the D nucleus splits the transition into three components,
labeled by the total angular momentum $F = J + I$.
The three expected hyperfine components are clearly detected and resolved.
We derive the frequencies of these components by fitting three independent
Gaussians to the observed spectrum. The fit uses a cloud velocity of
\vlsr\,=\,7.069~\kms\ (see \S~\ref{s:obs}) and a line width of 35\,kHz which is
the average width of the three components.
The spacings of the components, independent of \vlsr, are $-$0.198~(6) and
+0.269~(6)~\kms\ which is +47.58~(144) and $-$64.64~(144)~kHz
(Table~\ref{t:dcop-data}). These splittings are linear combinations of the
quadrupole coupling constant $eQq$ and the spin-rotation constant $C_I$ of the D
nucleus (e.g., \citealt{townes:bible}).  Solving the two equations with two
unknowns gives the $eQq$ and $C_I$ values given in Table~\ref{t:dcop-par}.

%%%%%%%%%%%%%%%%%%%%%%%%%%%%%%%%%%%%%%%%%%%%%%%%%%%%%%%%%%%%%%%%%%%%%%%%%%%%%%
\begin{table*}[p]
  \caption{Measured hyperfine splitting $\varDelta ^a$ (kHz) and residual o$-$c 
    of \dcop, DNC, and HN\thc\ relative to a selected hyperfine component. If no
    numbers are given for the value of a measured splitting and its residual,
    the hyperfine component overlaps with the previous component. The values of
    the splitting and the residual refer to the respective intensity weighted
    averages.} 
\label{t:dcop-data}

\begin{tabular}{lr@{}lr@{}l}
\hline \hline
Transition      & \multicolumn{2}{c}{Splitting} & \multicolumn{2}{l}{Residual$^b$} \\
\hline
\multicolumn{5}{l}{\dcop: $F = 2 - 1$ at 72039.306~(3) MHz} \\
\hline
$1 - 1$         &   +47&.58~(144)&    1&.20 \\
$0 - 1$         & $-$64&.64~(144)&    0&.80 \\
\hline
\multicolumn{5}{l}{DNC: $I,F = 2,1 - 2,2 / 2,1 - 0,0$ at 76305.512~(3) MHz} \\
\hline
$1,1 - 1,1$     &  115&.30~(100) & $-$0&.18 \\
$2,3 - 2,2$     &  175&.44~(100) & $-$0&.72 \\
$1,2 - 1,1$     &  204&.63~(100) &    1&.38 \\
$2,2 - 2,2$     &  277&.57~(100) & $-$0&.82 \\
$0,1 - 0,0$     &     &          &     &    \\       
$0,1 - 2,2$     &     &          &     &    \\
$1,0 - 1,1$     &  325&.50~(250) &    1&.60 \\
\hline
\multicolumn{5}{l}{HN\thc: $F_1,F_2,F = 0,1,1 - 1,2,2$ at 87090.675~(3) MHz} \\
\hline
$2,2,2 - 1,1,1$ &  115&.5~(50)   & $-$7&.3 \\
$2,2,2 - 1,2,2$ &     &          &     &   \\
$2,2,1 - 1,1,0$ &     &          &     &   \\
$2,2,1 - 1,2,1$ &     &          &     &   \\
$2,3,3 - 1,2,2$ &  158&.5~(50)   &    5&.0 \\
$2,3,2 - 1,1,1$ &     &          &     &   \\
$2,3,2 - 1,2,1$ &     &          &     &   \\
$1,1,1 - 1,1,1$ &  210&.7~(50)   &    1&.5 \\
$1,2,2 - 1,2,1$ &     &          &     &   \\
$1,2,2 - 1,2,2$ &     &          &     &   \\

\hline

\end{tabular}
\medskip

$^a$ Numbers in parentheses are one standard deviation in units 
 of the least significant figures.\\
$^b$ Residuals for DCO$^+$ refer to fit with $C_I$ kept fixed. 
 The residuals are zero if $C_I$ is released as two pieces of information 
 are then used to derive two parameters.\\
 
\end{table*}
%%%%%%%%%%%%%%%%%%%%%%%%%%%%%%%%%%%%%%%%%%%%%%%%%%%%%%%%%%%%%%%%%%%%%%%%%%%%%%

%%%%%%%%%%%%%%%%%%%%%%%%%%%%%%%%%%%%%%%%%%%%%%%%%%%%%%%%%%%%%%%%%%%%%%%%%%%%%%
\begin{table*}[p]
\caption{Derived hyperfine parameters$^a$ (kHz) of \dcop\ in comparison to
  previous experimental and present quantum-chemically calculated values.}
\label{t:dcop-par}

\begin{tabular}{lr@{}lr@{}lr@{}lr@{}l}

\hline \hline

Parameter   & \multicolumn{8}{c}{Value} \\
\cline{2-9}
            & \multicolumn{2}{c}{This work} & \multicolumn{2}{c}{This work} & \multicolumn{2}{c}{\citet{caselli:dco+}} & \multicolumn{2}{c}{calculated} \\
\hline

$eQq$         & +150&.00~(266) & +151&.12~(288) & +147&.8~(35)  & +156&.0  \\
$C_I$         & $-$0&.69$^b$   & $-$1&.12~(43)  & $-$1&.59~(78) & $-$0&.69 \\

\hline
\end{tabular}

\medskip

$^a$ Numbers in parentheses are one standard deviation in units 
of the least significant figures.\\
$^b$ Kept fixed in the analysis.\\

\end{table*}
%%%%%%%%%%%%%%%%%%%%%%%%%%%%%%%%%%%%%%%%%%%%%%%%%%%%%%%%%%%%%%%%%%%%%%%%%%%%%%

The present value of $C_I$ in Table~\ref{t:dcop-par} is smaller in magnitude
than that by \citet{caselli:dco+} while $eQq$ is larger. However, agreement
exists within the larger uncertainties from \citet{caselli:dco+}.  These authors
also observed the LDN 1512 cloud with the IRAM 30m telescope, but at a different
position, where the lines are weaker.  Therefore, the signal-to-noise ratio of
their spectra is lower, even though the integration time has been longer.

The quantum-chemical calculations for \dcop\ (see Table \ref{t:qcc-theor}) yield
as best values for $eQq$ 156.0 kHz (CCSD(T)/cc-pCV5Z calculations plus
vibrational corrections); the small changes from the quadruple-zeta (QZ) to the
quintuple-zeta (5Z) basis set suggests a minute increase of probably less than 1~kHz
upon extrapolation to the basis-set limit.  The agreement is good, within twice
the experimental uncertainties, for the fit in which $C_I$ has been varied.  The
agreement is slightly worse for the fit in which $C_I$ was kept fixed to the
quantum-chemically calculated value and slightly worse still in comparison to
the previous values from \citet{caselli:dco+}, see Table~\ref{t:dcop-par}.

The best theoretical value for $C_I$ is $-$0.69~kHz. As discussed in
\S~\ref{s:qcc}, no vibrational corrections could be computed for this
value. However, since the vibrational correction to $eQq$ is fairly small, it is
likely that the one on $C_I$ will be small also, possibly of order of 0.1~kHz.
The convergence patterns seen in Table~\ref{t:nsr-theor} suggest that remaining
errors in the calculations can be considered rather small so that we conclude,
based on these theoretical estimates as well as the results of our measurements
that \citet{caselli:dco+} overestimated $C_I$ due to the limited signal-to-noise
in their spectrum. Both fitted values agree within the fairly large
uncertainties with the theoretical value.

The absolute positions of the strongest hyperfine component from the present 
astronomical observations agree to about 3~kHz with the previous observation
\citep{caselli:dco+}.

%%%%%%%%%%%%%%%%%%%%%%%%%%%%%%%%%%%%%%%%%%%%%%%%%%%%%%%%%%%%%%%%%%%%%%%%%%%%%%
\subsection{Anomalous excitation of DCO$^+$}
\label{sec:3j}
%%%%%%%%%%%%%%%%%%%%%%%%%%%%%%%%%%%%%%%%%%%%%%%%%%%%%%%%%%%%%%%%%%%%%%%%%%%%%%

The continuous line in Figure~\ref{f:dcop} is a model fit to the data that uses
the splitting constants that we just derived.  The model was computed with the
hyperfine structure (HFS) routine within CLASS. This routine assumes that the
hyperfine components of the transition have the same width and the same
excitation temperature. While the data support the first assumption, the second
apparently does not hold. The highest-frequency $F = 1 - 1$ component has a
higher intensity than the model predicts.  Such `hyperfine anomalies' are
well-known for the $J = 1 - 0$ line of HCN, where the $F = 0 - 1$ component is
unexpectedly strong toward many dark clouds (e.g., \citealt{walmsley:hcn}). The
implied difference in excitation temperatures between the hyperfine components
can be understood through detailed calculations of hyperfine selective
collisional cross sections \citep{monteiro}.  Such calculations do not exist for
\dcop\ at present, although they have been reported for \nnhp\ \citep{daniel05},
where a similar excitation anomaly has been observed \citep{caselli:n2h+}.
Hyperfine selective collision data for the electron impact excitation of HCN,
HNC, DCN and DNC are presented by \citet{faure:hcn-hfs}, but do not apply here,
since the electron abundance in dark clouds is low.

%%%%%%%%%%%%%%%%%%%%%%%%%%%%%%%%%%%%%%%%%%%%%%%%%%%%%%%%%%%%%%%%%%%%%%%%%%%%%%
\subsection{Hyperfine splitting of DNC}
\label{sec:dnc}
%%%%%%%%%%%%%%%%%%%%%%%%%%%%%%%%%%%%%%%%%%%%%%%%%%%%%%%%%%%%%%%%%%%%%%%%%%%%%%

%%%%%%%%%%%%%%%%%%%%%%%%%%%%%%%%%%%%%%%%%%%%%%%%%%%%%%%%%%%%%%%%%%%%%%%%%%%%%%
\begin{figure*}[p]
\includegraphics[width=7cm,angle=0]{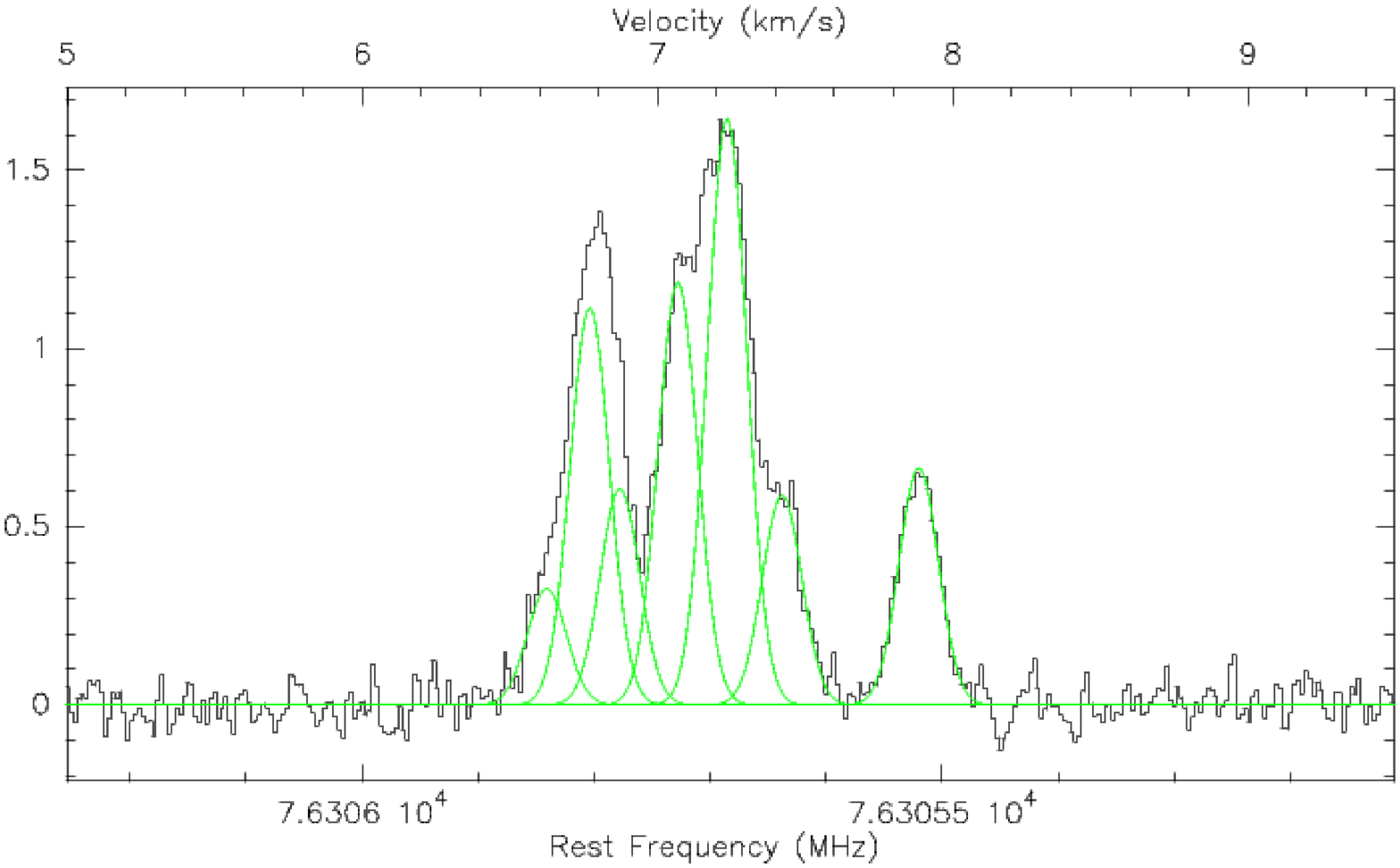}
\includegraphics[width=7cm,angle=0]{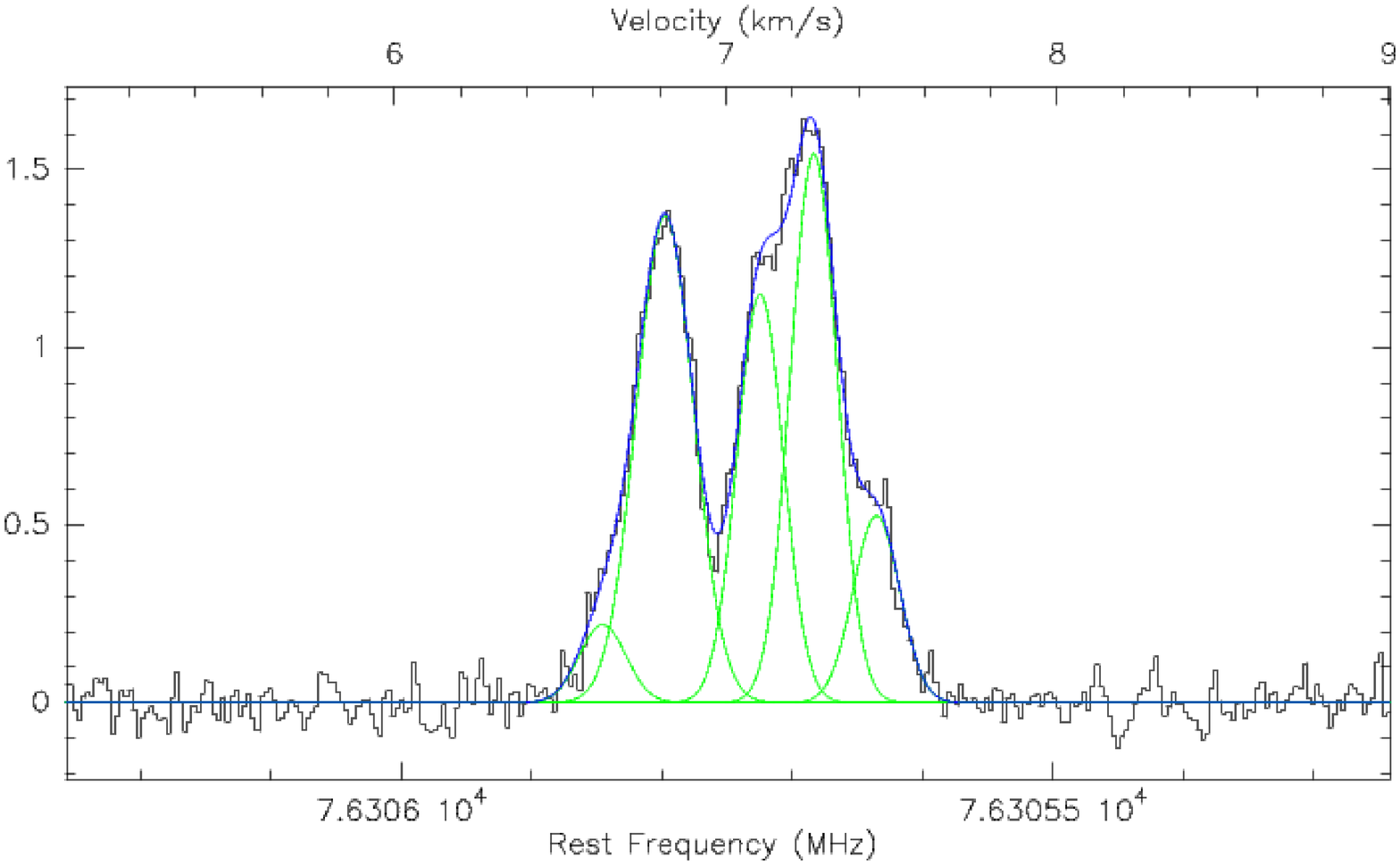}
\caption{Spectrum of the $J = 1 - 0$ transition of DNC toward the dark cloud LDN
  1512, obtained with the IRAM 30m telescope. Vertical scale is antenna
  temperature ($T_{\rm mb}$) in K. \underline{Left panel:} Gaussian
  decomposition of the broad blended feature into six components which are
  assumed to have the same width as the single isolated
  feature. \underline{Right panel:} Spectrum with the isolated feature
  subtracted and a reduced five-component fit to the remaining broad feature. }
\label{f:dnc}
\end{figure*}
%%%%%%%%%%%%%%%%%%%%%%%%%%%%%%%%%%%%%%%%%%%%%%%%%%%%%%%%%%%%%%%%%%%%%%%%%%%%%%

Figure~\ref{f:dnc} shows the spectrum of DNC toward LDN 1512.  The profile
consists of one weak, isolated component at low frequency, and two stronger,
broader peaks which appear to be blends of several components.  The spins $I=1$
of the D and $^{14}$N nuclei lead theoretically to a splitting into nine
components at seven different frequencies, but not all these components are
resolved in the current data. Since the hyperfine splitting of both nuclei are
of similar magnitude, we have used the symmetric coupling scheme $I_{\rm tot} =
I_{\rm N} + I_{\rm D}$ and $F = J + I_{\rm tot}$.  In contrast,
\citet{bechtel:hnc-dnc} use the sequential coupling scheme, which only affects
the labeling of the levels, but not the shape of the hfs pattern.  The isolated
component is well fitted with a Gaussian profile of width 0.150~(1)~\kms.  We
have fitted the profile with an `empirical' model consisting of several
Gaussians with independent positions and intensities, but with the width fixed
to 0.15~\kms.  The fit has significant residuals if four or five Gaussians are
used, but the sum of six Gaussians (plus one for the isolated feature) gives a
satisfactory fit to the observed profile (Fig.~\ref{f:dnc}a).

The results of the empirical model are in good agreement with our quantum-chemical
calculations, which predict seven hyperfine components for this transition, 
due to the nuclear spin of both D and N.  In the case of DNC, the quadrupole 
moments $eQq$ of the D and N nuclei are similar in magnitude, 
unlike in DCN where $eQq({\rm N}) \gg eQq({\rm D})$.

The positions of the components of the empirical model are in good agreement
with the predictions based on quantum-chemical calculations, except for the
highest-frequency component.  Suspecting that this mismatch is caused by the
small separation of the 2,2$\to$2,2 and the (coinciding) 0,1$\to$0,0 and
0,1$\to$2,2 components, we have reduced the empirical model to five Gaussians,
and freed up the width of the fourth Gaussian, which represents these
components.  This `semi-empirical' model gives an excellent match to the data,
as shown by the solid line in Fig.~\ref{f:dnc}b. The width of the fourth
Gaussian is 0.202~(9)~\kms\ implies a separation of the 2,2$\to$2,2 component
from the 0,1$\to$0,0 and 0,1$\to$2,2 components of 0.05~\kms. This result is in
good agreement with the theoretical results: the calculated splitting is
8.7\,kHz, but the apparent separation is smaller because of the small intensity
of the 0,1$\to$0,0 and 0,1$\to$2,2 components; see Table~\ref{t:dcop-data}.

%%%%%%%%%%%%%%%%%%%%%%%%%%%%%%%%%%%%%%%%%%%%%%%%%%%%%%%%%%%%%%%%%%%%%%%%%%%%%%
\begin{table*}[p]
\caption{Hyperfine parameters$^a$ (kHz) of DNC in comparison to previous 
         experimental and present quantum-chemically calculated values.}
\label{t:dnc}

\begin{tabular}{lr@{}lr@{}lr@{}l}
\hline 
\hline
Parameter   & \multicolumn{6}{c}{Value} \\
\cline{2-7}
            & \multicolumn{2}{c}{This work} & 
   \multicolumn{2}{c}{\citet{bechtel:hnc-dnc}} & \multicolumn{2}{c}{ab initio} \\
\hline

$eQq$(N) & +288&.2~(71)  & +294&.7~(131) & +309&.6  \\
$C_I$(N) &   +4&.91~(63) &   +5&.01~(99) &   +5&.54 \\
$eQq$(D) & +265&.9~(83)  & +261&.9~(145) & +257&.6  \\
$C_I$(D) & $-$0&.60$^b$  &     &$-$      & $-$0&.60 \\
$S$(ND)  & $-$1&.35$^b$  &     &$-$      &     &$-$ \\

\hline
\end{tabular}

\medskip

$^a$ Numbers in parentheses are one standard deviation in units 
of the least significant figures.\\
$^b$ Kept fixed to computed values in the analysis.\\
\end{table*}
%%%%%%%%%%%%%%%%%%%%%%%%%%%%%%%%%%%%%%%%%%%%%%%%%%%%%%%%%%%%%%%%%%%%%%%%%%%%%%

The hyperfine parameters determined for DNC are compared in Table~\ref{t:dnc}
with those previously determined in the laboratory by \citet{bechtel:hnc-dnc}
and with the best ab initio values.  We note a good agreement between the two
experimental sets of data, but emphasize that our values have reduced error
bars.  The comparison with the theoretical best estimates is also favorable and
lends support to the reliability of the experimentally determined hyperfine
parameters.  The theoretical quadrupole coupling terms determined with the
largest basis set and incorporating vibrational corrections are 257.6 and
309.6~kHz for the D and $^{14}$N nucleus, respectively, while the calculated
nitrogen spin-rotation constant amounts to 5.54~kHz. The theoretical values are
apparently slightly closer to the previous values \citep{bechtel:hnc-dnc}.
While a smooth and fast convergence is observed in the calculation of the
spin-rotation and deuterium quadrupole coupling constants, the determination of
the nitrogen quadrupole coupling constant turned out to be challenging. Going
from the quadruple-zeta to the quintuple-zeta set causes a change of 16 kHz and
vibrational corrections amount to more than 40 kHz, while electron-correlation
effects beyond CCSD(T) are found to be less important.  However, the noted
problems in the accurate determination of the nitrogen quadrupole coupling
constant can be traced back to the smallness of the actual value which is
roughly one order of magnitude smaller than the corresponding quadrupole
coupling in DCN \citep{Lamb-dip_DCN}. Moreover, the changes in the calculated
$eQq$ values decrease fairly rapidly by a factor of about 6 with increasing size
of the basis set. Taking into account the decrease in the magnitude of the
vibrational corrections one can estimate a value of about $306.5 \pm 1$~kHz as
the CCSD(T) value at the inifinitely large basis set, in agreement within the
uncertainty of the laboratory value and within 2.5 times the uncertainties of
the present astronomical observations.

The absolute positions of the isolated hyperfine components from the present 
astronomical observations agree to 1~kHz with the previous laboratory frequencies
\citep{bechtel:hnc-dnc}. The laboratory frequency was deemed to be uncertain 
to 5~kHz while our astronomical frequency is uncertain to probably no more than 3~kHz.

%%%%%%%%%%%%%%%%%%%%%%%%%%%%%%%%%%%%%%%%%%%%%%%%%%%%%%%%%%%%%%%%%%%%%%%%%%%%%%
\subsection{Hyperfine splitting of HN\thc}
\label{sec:hn13c}
%%%%%%%%%%%%%%%%%%%%%%%%%%%%%%%%%%%%%%%%%%%%%%%%%%%%%%%%%%%%%%%%%%%%%%%%%%%%%%

\begin{figure}[p]
\includegraphics[width=7cm,angle=0]{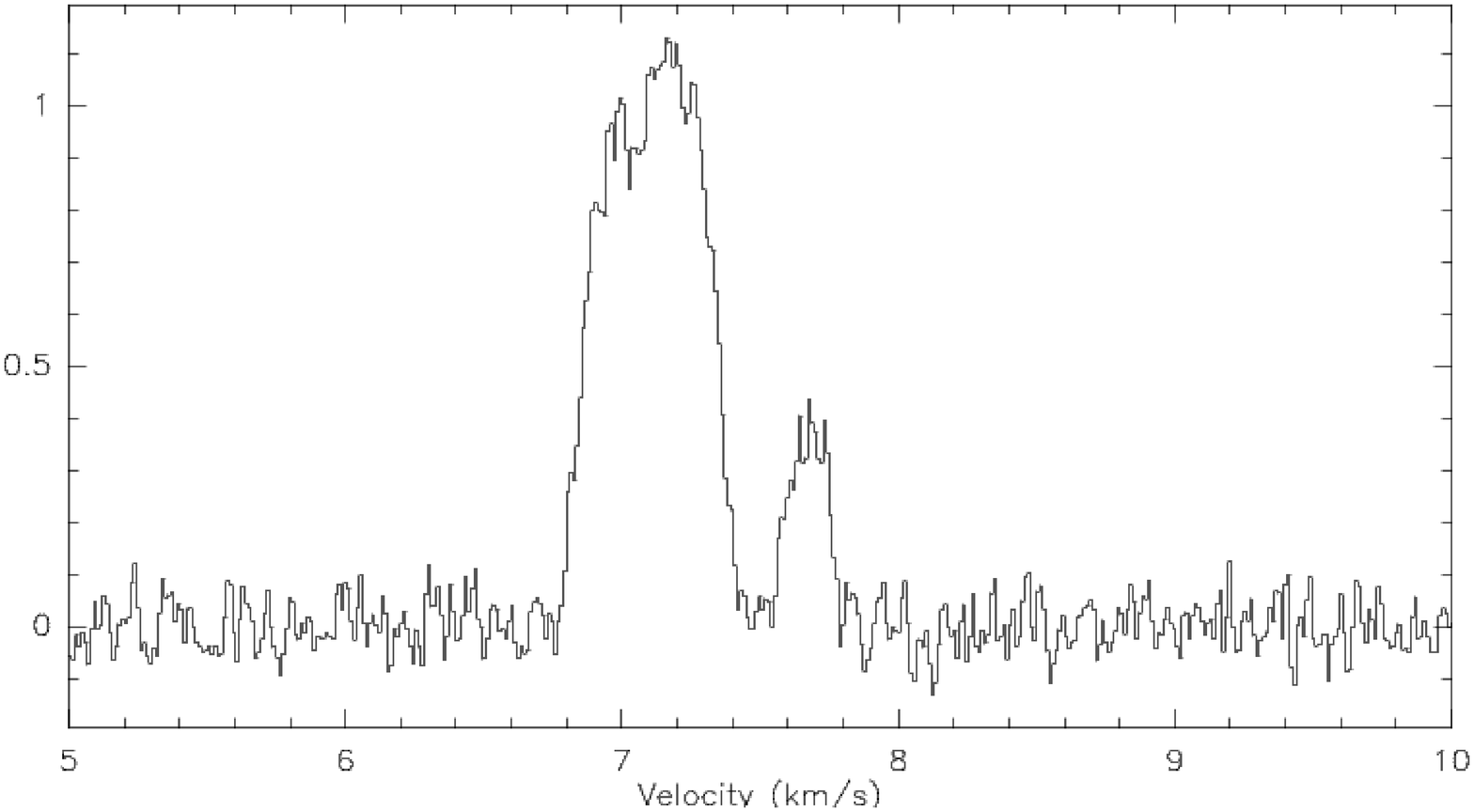}
\caption{Spectrum of the $J = 1 - 0$ transition of HN\thc\ toward the dark cloud
  LDN 1512, obtained with the IRAM 30m telescope. Vertical scale is antenna
  temperature ($T_{\rm mb}$) in K.}
\label{f:hn13c}
\end{figure}

Figure~\ref{f:hn13c} shows the spectrum of HN\thc\ $J = 1 - 0$ toward LDN~1512.
One weak, isolated hyperfine feature at low frequency (LF for short) as well as
two somewhat resolved ones can be recognized easily as would be expected for a
molecule with a $^{14}$N nucleus.  Analyses of the astronomical spectrum as well
as simulations of the spectrum based on the experimental quadrupole moment
($eQq$) for HNC \citep{bechtel:hnc-dnc}, the quantum-chemically calculated
nuclear spin-rotation parameters ($C_I$) and the spin-spin coupling terms
($S_{XY}$) derived from the interatomic distances and the nuclear magnetic
moments reveal that the situation is more complicated. First, contributions from
the \thc\ nucleus considerably broaden the central hyperfine feature (CF for
short) such that a shoulder to the lower frequency side should be discernable.
Second, the hyperfine pattern should get wider, and the relative intensity of
the highest frequency feature (HF for short) should increase somewhat with
respect to the central one. Splitting caused by the H nucleus increases the
total number of hyperfine components further, but the total number of observable
features does not increase.  However, it is noteworthy that the H hyperfine
parameters counteract somewhat the effects caused by the \thc\ nucleus mentioned
under "second".  The net effect is one isolated, weak hyperfine feature at low
frequencies, one strong and very broad CF with a low-frequency shoulder, and a
HF that should be stronger than the isolated component, but weaker than the
central one.

Analyses were started assuming three different hyperfine features with three
different widths. The LF-HF splitting seemed reasonable, but the HF was too
strong compared with the CF, and the position of the CF was too low in
frequency.  Fitting two features for the CF required their width to be equal to
that of the LF which was determined to be 43\,kHz as expected from the
simulations.  A fit of four components with widths of 43\,kHz was still fairly
poor.  Simulations suggested the width of the HF to be about 15\,\% larger than
that of the remaining components. The analysis of the astronomical spectrum
improved somewhat, but was still not satisfactory. Moreover, a fit of $eQq$ to
the splittings with all other hyperfine parameters kept fixed yielded a value of
$272.5 \pm 5.1$~kHz, which is larger than the one determined for HNC, $264.5 \pm
4.6$~kHz, in the laboratory \citep{bechtel:hnc-dnc}, but in agreement within the
quoted uncertainties.  The rms of the fit is 5.1~kHz.  Releasing the width of
the HF gave the best analysis of the astronomical spectrum, but yielded a width
of $71.5 \pm 2.3$~kHz, much bigger than expected from the simulations. The
spectroscopic fit yields $eQq = 258.7 \pm 5.1$~kHz, which is now smaller than
the HNC value, but again in agreement within uncertainties. The rms of the fit
is, with 4.0~kHz, slightly better than the fit mentioned before (Table~\ref{t:dcop-data}).

The contributions to the hyperfine splittings decrease from $^{14}$N 
over \thc\ to H. Therefore, we apply a sequential coupling scheme:
$F_1 = J + I$($^{14}$N); $F_2 = J + I$($^{13}$C); $F = J + I$(H). 
Only $eQq$ could be determined for the $^{14}$N nucleus. Trial fits 
with $C$($^{14}$N) or $C$($^{13}$C) or both released yielded values for 
these parameters that were deemed to be unreliable 
as they could differ by as much as about 5~kHz from the predicted values.

%%%%%%%%%%%%%%%%%%%%%%%%%%%%%%%%%%%%%%%%%%%%%%%%%%%%%%%%%%%%%%%%%%%%%%%%%%%%%%
%%%%%%%%%%%%%%%%%%%%%%%%%%%%%%%%%%%%%%%%%%%%%%%%%%%%%%%%%%%%%%%%%%%%%%%%%%%%%%

%%%%%%%%%%%%%%%%%%%%%%%%%%%%%%%%%%%%%%%%%%%%%%%%%%%%%%%%%%%%%%%%%%%%%%%%%%%%%%
\begin{table*}[p]
\caption{Hyperfine parameters$^a$ (kHz) of HN\thc\ in comparison to ab initio values.}
\label{t:hnc-13}

\begin{tabular}{lr@{}lr@{}l}

\hline
\hline

Parameter   & \multicolumn{4}{c}{Value} \\
\cline{2-5}
            & \multicolumn{2}{c}{experimental} & \multicolumn{2}{c}{ab initio} \\
\hline

$eQq$(N)     & +272&.5~(51) & +279&.5  \\
$C_I$(N)     &   +6&.33$^b$ &   +6&.33 \\
$C_I$(\thc\ )&  +23&.46$^b$ &  +23&.46 \\
$C_I$(H)     & $-$4&.49$^b$ & $-$4&.49 \\
$S$(HN)      & $-$8&.79$^b$ &     &$-$ \\
$S$(N\thc\ ) & $-$1&.37$^b$ &     &$-$ \\
$S$(H\thc\ ) & $-$2&.98$^b$ &     &$-$ \\

\hline
\end{tabular}

\medskip

$^a$ Numbers in parentheses are one standard deviation in units 
of the least significant figures.\\
$^b$ Kept fixed to computed values in the analysis.\\

\end{table*}
%%%%%%%%%%%%%%%%%%%%%%%%%%%%%%%%%%%%%%%%%%%%%%%%%%%%%%%%%%%%%%%%%%%%%%%%%%%%%%

The final values for the hyperfine parameters for HN$^{13}$C are given
in Table \ref{t:hnc-13}. The resulting nitrogen quadrupole coupling 
of 272.5~(51) kHz is in good agreement with the best theoretical
value of 279.5 kHz, in particular when noting the slow basis-set 
convergence in the corresponding calculations. The vibrational contributions 
are much larger still than the already large contributions for DNC; 
they amount to about 46~kHz for HN$^{13}$C. These contributions decrease 
again slightly in magnitude going from the quadruple-zeta to the quintuple-zeta 
basis set. 
Thus, the CCSD(T) $eQq$($^{14}$N) value at the basis-set limit is probably 
around 3~kHz smaller than the one calculated at the cc-pCV5Z basis set 
with vibrational corrections and now agreeing within the uncertainty 
with the value from astronomical observations.

%%%%%%%%%%%%%%%%%%%%%%%%%%%%%%%%%%%%%%%%%%%%%%%%%%%%%%%%%%%%%%%%%%%%%%%%%%%%%%
\subsection{Rotational and distortion parameters for HN\thc}
\label{B_D_hn13c}
%%%%%%%%%%%%%%%%%%%%%%%%%%%%%%%%%%%%%%%%%%%%%%%%%%%%%%%%%%%%%%%%%%%%%%%%%%%%%%

Accurate rotational as well as distortion parameters are available for DNC
\citep{DNC_2THz,bechtel:hnc-dnc} and DCO$^+$
\citep{caselli:dco+,HCO+-isos_rot_2007} and the present measurements will only
modify these parameters slightly.  In contrast, only sparse data are available
for HN\thc.  \citet{HNC-isos_rot_1976} measured the $J$=1$\gets$0 and 2$\gets$1
rotational transitions of several HNC isotopologs. More recently,
\citet{HNC_IR_2001} obtained rovibrational transitions of the $\nu _2$ bending
mode of HN\thc\ together with extensive data for HNC. Replacing the
$J$=1$\gets$0 laboratory rest frequency with our astronomical rest frequency
(Table~\ref{t:dcop-data}) reduces the uncertainties of $B$ and $D$ by
factors of 5 and 1.7, respectively.  Fixing $H$ to a value of 157~mHz taken from
HNC \citep{HNC_2THz,bechtel:hnc-dnc}, we obtain $B = 43545.6000~(47)$~MHz and $D
= 93.7~(20)$kHz; the numbers in parentheses are one standard deviation in units
of the least significant figures.
The present values of $B$ and $D$ agree with the previous ones within their
larger uncertainties \citep{HNC-isos_rot_1976}.  The resulting predictions are
accurate up to $\approx$500\,GHz but should be viewed with some caution at
higher frequencies.

%%%%%%%%%%%%%%%%%%%%%%%%%%%%%%%%%%%%%%%%%%%%%%%%%%%%%%%%%%%%%%%%%%%%%%%%%%%%%%
\section{ Discussion and conclusions}
\label{s:disc}
%%%%%%%%%%%%%%%%%%%%%%%%%%%%%%%%%%%%%%%%%%%%%%%%%%%%%%%%%%%%%%%%%%%%%%%%%%%%%%

\subsection{Astronomical implications}
\label{sec:astro-impl}

At spectral resolutions of $\sim$0.1\,\kms, as commonly achieved by astronomical
instrumentation in the 3\,mm wavelength range, the $J=1$--0 line of \dcop\
splits into three hyperfine components, the HN\thc\ $J=1$--0 line into four, and
the DNC $J=1$--0 line into six. Table~\ref{t:astro} gives the velocity offsets and
the relative strengths of these components. The strengths are theoretical values
which only apply if the lines are optically thin and the excitation is
thermalized; as Section~\ref{sec:3j} discusses, these conditions are often not
fulfilled, so that detailed radiative transfer modeling is needed to extract
physical information out of the observed hyperfine intensities.

The hyperfine splitting extends over 0.47\,\kms\ for \dcop, over 1.28\,\kms\ for
DNC, and over 0.73\,\kms\ for HN\thc. These intervals are comparable to or
larger than the total (thermal + turbulent) line widths in many astrophysical
objects, in particular pre- and protostellar cores in star-forming regions
\citep{bergin:araa}. Taking these splittings into account is therefore crucial
to derive accurate column densities of \dcop, DNC and HN\thc\ from their
ground-state rotational lines. 

The magnitude of the hyperfine splittings is also comparable to the velocities
of infalling, outflowing and rotational motions around protostars on
$\sim$1000\,AU scales (e.g., \citealt{hogerheijde01}). Taking the splitting into
account is therefore essential for a correct determination of the velocity field
of the gas from these lines. The quadrupole splittings of linear molecules
  decrease rapidly with increasing $J$, at least for the strong transitions, so
  that going to higher-$J$ lines would seem a way to avoid this complication.
However, at the typical (low)
temperatures and densities of pre- and protostellar cores, the higher-$J$ lines
of \dcop, DNC and HN\thc\ are usually less excited and thus weaker than the
ground-state lines.

\begin{table*}[p]
\caption{Velocity offsets and intrinsic strengths of hyperfine components of the
\dcop, DNC, and HN\thc\ $J=1$--0 lines.}
\label{t:astro}

\begin{tabular}{llll}

\hline
\hline

Transition       & Component                                           & Offset & Relative \\
                 &                                                     & \kms\  & intensity \\
\hline

\dcop\ $J=1$--0  & $F$=0--1                                               & $+$0.269 & 0.20 \\
                 & $F$=2--1                                               & $\pm$0.0 & 1.00 \\
                 & $F$=1--1                                               & $-$0.198 & 0.60 \\
DNC $J=1$--0     & $I,F$=2,1--2,2 / 2,1--0,0                              & $\pm$0.0 & 1.00 \\
                 & $I,F$=1,1--1,1                                         & $-$0.453 & 0.99 \\
                 & $I,F$=2,3--2,2                                         & $-$0.689 & 2.33 \\
                 & $I,F$=1,2--1,1                                         & $-$0.804 & 1.67 \\
      & $I,F$=2,2--2,2 / 0,1--0,0 / 0,1--2,2                              & $-$1.091 & 2.67 \\
                 & $I,F$=1,0--1,1                                         & $-$1.279 & 0.33 \\
HN\thc\ $J=1$--0 & $F_1,F_2,F$=0,1,1--1,2,2                               & $\pm$0.0 & 1.00 \\
  & $F_1,F_2,F$=2,2,2--1,1,1 / 2,2,2--1,2,2 / 2,2,1--1,1,0 / 2,2,1--1,1,0 & $-$0.398 & 4.04 \\
  & $F_1,F_2,F$=2,3,3--1,2,2 / 2,3,2--1,1,1 / 2,3,2--1,2,1                & $-$0.546 & 6.63 \\
  & $F_1,F_2,F$=1,1,1--1,1,1 / 1,2,2--1,2,1 / 1,2,2--1,2,2                & $-$0.725 & 3.66 \\

\hline
\end{tabular}

\medskip

%$^a$ \\

\end{table*}

\subsection{Molecular physics implications}
\label{sec:molphys-impl}

The most remarkable aspect of the present results from the molecular physics
point of view are the large vibrational corrections to the $^{14}$N $eQq$ values
of HN$^{13}$C and DNC and the corresponding large differences in the ground
state $eQq$ values. Isotopic differences in experimentally determined quadrupole
coupling constants are usually not or barely significant.  Significant effects
have been observed e.g. by \citet{XBE_1998} where $^{10}$B/$^{11}$B isotopic
shifts of around 15~kHz were determined for quadrupole coupling constants of
both $^{35}$Cl and $^{37}$Cl of ClBO, though these shifts were only very small
fractions ($< 10^{-3}$) of the constants themselves. The authors attributed
these effects to the non-rigidity of the molecule in particular along the
bending coordinate.  Experimental work \citep{XCN_XNC_vib_exp_2008} and
calculations \citep{XCN_XNC_vib_ai_2008} lend support to this view and suggest
that low-order vibrational corrections to the $eQq$ values are probably not
sufficient for very accurate predictions of quadrupole coupling constants for
specific vibrational states.

While the bending mode of  HNC is known to be fairly non-rigid, those of the 
isoelectronic HCN, HCO$^+$, and N$_2$H$^+$ molecules are much more rigid. 
However, the bending mode of the HOC$^+$ molecule is even less rigid than 
that of HNC. Therefore, it would be interesting to investigate the $^{17}$O 
quadrupole coupling constants of H$^{17}$OC$^+$ and D$^{17}$OC$^+$. 
The HOC$^+$ molecule is less abundant than its isomer HCO$^+$ by factors of $\sim$100
in diffuse clouds and $\sim$1000 in dense clouds \citep{liszt04}, but HOC$^+$ is
strongly enhanced in regions with strong radiation fields and/or shocks
\citep{rizzo03,fuente05,fuente08}, so these may be good places to search for its
rare isotopologues.
In addition, this molecule is not so easy to produce in the laboratory. 
Thus, the easiest way to obtain information on the hyperfine structure of 
this ion are detailed quantum-chemical calculations.

New or updated predictions of the rotational spectra for DCO$^+$, DNC, and
HN\thc\ will be available in the catalog section of the Cologne Database for
Molecular spectroscopy\footnote{website: http://www.astro.uni-koeln.de/cdms/}
\citep{CDMS_1,CDMS_2}.

%%%%%%%%%%%%%%%%%%%%%%%%%%%%%%%%%%%%%%%%%%%%%%%%%%%%%%%%%%%%%%%%%%%%%%%%%%%%%%
\begin{acknowledgement}
  The authors thank Michael Grewing (IRAM) for awarding
  Director's Time to our observing project, Johannes Schmid-Burgk (MPIfR) for
  discussions about the LDN 1512 cloud, Gabriel Paubert (IRAM) for discussions
  on heliocentric corrections, and Arnaud Belloche and Dirk Muders (MPIfR) for
  assistance with the observations.  H.S.P.M. thanks the Deutsche
  Forschungsgemeinschaft (DFG) for initial support through the collaborative
  research grant SFB~494. He is grateful to the Bundesministerium f\"ur Bildung
  und Forschung (BMBF) for recent support which was administered by the
  Deutsches Zentrum f\"ur Luft- und Raumfahrt (DLR). M.E.H. and J.G. acknowledge
  support from the Deutsche Forschungsgemeinschaft (DFG) and the 
  Fonds der Chemischen Industrie.
\end{acknowledgement}
%%%%%%%%%%%%%%%%%%%%%%%%%%%%%%%%%%%%%%%%%%%%%%%%%%%%%%%%%%%%%%%%%%%%%%%%%%%%%%

\bibliographystyle{aa}
\bibliography{hfs,spec}

\end{document}